\documentstyle[12pt]{article}

         \def\ee{\end{equation}}
         \def\be{\begin{equation}}
         \def\ba{\begin{array}}
         \def\ea{\end{array}}
         \def\bea{\begin{eqnarray}}

\begin{document}

\vskip2cm

\begin{center}

\vskip2cm

\begin{huge}
{\bf Static Parameters of Hadrons and Quantum Groups}
\end{huge}
\vskip 1.0cm
{\bf F. Kazemi Tabatabaei}
\footnote{e-mail : fariba@netware2.ipm.ac.ir} \\
\vspace{0.50cm}
 {\it Institute for Studies in Theoretical Physics and Mathematics(IPM)  \\
P.O. Box.: 19395-5746, Tehran, Iran } \\
\vspace{0.50cm}
 {\it Dept. of Particle Physics, University of Santiago de Compostela, \\
Santiago de Compostela, Spain}
\end{center}
\vspace{1.0cm}
\begin{abstract}
We study the static properties of hadrons, assuming quantum group symmetry.
We calculate the magnetic moment, axial form factor and A-symmetry, using
$SU_q(2)$ and $SU_q(3)$ quantum groups. The results are fitted with
experimental data, giving an interval of $0.9<q<1.1$.
Some of the implications for the deformation parameter are discussed.

\end{abstract}
\newpage

\section{Introduction}
Despite rather extended study of quantum groups as mathematical objects[1-7],
little has been done in the way of connecting them with observable phenomena,
but some papers in this direction have appeared [8-11].
In such theories, the deformation parameter q is a function of
a dimensionless combination of constants, so that in the limit of
$ q \rightarrow 1$, we recover the undeformed theories, this may for example
correspond to the low energy limit. For instance, Newtonian mechanics
can be thought of as the low velocity limit ( $ v/c \rightarrow 0 $) of
the relativistic mechanics. In this paper we use quantum groups in
deforming the flavour symmetry of hadrons,  with the help of the standard
q-deformation \cite{D} of $SU_q(2)$ and $SU_q(3)$.
Knowing that flavour symmetry is not an exact symmetry of nature,
we chose to deform
the flavour group rather than the other symmetry groups of hadrons. However,
 as we shall see later, it becomes clear that deformation of either
 spin or colour symmetry is unavoidable.
 We placed hadrons in representations of quantum
 groups so that in the
limit $ q \rightarrow 1$, the standard decuplet, octet and singlet
representations are recoverd.

The price of using quantum groups as symmetry group of fundamental particles
is to abandon the notion of permutation symmetry of fermions and bosons.
The representation theory of the classical groups is closely linked with
the group of permutation of $n$ objects. Therefore, the concepts of
anti-symmetry of wavefunctions under the permutation of pairs of fermions
and symmetry under the permutation of pairs of bosons,can be incorporated
naturally, using anti-symmetric and symmetric tensor representations of
the classical groups. This device is not natural within the representation
theory of quantum groups. Here the permutation group is replaced by
its deformed version; the Braid group \cite{Y}. The braid group $B_n$ deals
with the  permutation of $n$ strings, and in general can be very complex.
The relevant braid group that we are concerned with in this paper,
 satisfies a quadratic relationship (Skein relation):
\be
B_{ij}^2+(q^2-1)B_{ij}=q^2
\ee
where $B_{ij}$, tangles two strands $i$ and $j$. This, replaces
the simple quadratic relationship of permuting two objects $i$ and $j$ in a
row of $n$ objects; $P_{ij}^2=1$. It is
clear that the relation (1.1) reduces to the permutation relation in the
limit $ q \rightarrow 1$. If we postulate that the physical states are
eigenstates of $B$ rather than $P$, then we will have no choice other
than letting bosons
correspond to an eigenstate with the unit eigenvalue and fermions to the
eigenstate with the eigenvalue equal to $-q^2$.
 Here we face a problem, that the octet states are not the
 eigenstates of the braid
matrix; the operation of $B$ on these states will not leave them invariant.
Thus the deformation of flavour alone
 can not be permitted, so we conclude that
 deforming another subspace such as spin is necessary \cite{WZ}.

The paper is organised as follows: in Sec.2 we give a brief introduction to
quantum groups, construct the q-deformed states of the flavour space, and
then together with  the q-deformed spin space, we
discuss the effects of the braid group.
In the Sec.3, following the standard procedures, we
calculate the static parameters of hadrons as functions of q, fitting
these calculations with the observed data, we find  an interval for q.
\section{Quantum Groups}
Every quantum group corresponds to a solution of the Yang-
Baxter equation, and to every solution we can correspond an integrable
statistical model \cite{B}. There are in fact mathematical relationships
between integrable models and quantum groups [1-2]. Here we shall only
give a brief discussion of the mathematical structure of the standard
deformation of $SU_q(3)$, which will suffice for our calculations.
The commutation relations of $SU_q(3)$ algebra are:
\be \ba{lll}
[H_i,H_j] &=& 0 \cr
[H_i,X_j^{\pm}] &=& a_{ij}X_j^{\pm}\cr [X_i^+,X_j^-] &=& \delta_{ij}
\frac{q^{H_i}-q^{-H_i}}{q-q^{-1}} \hskip2cm i,j=1,2
\ea \ee
where $H_i$, $X_i^{\pm}$ and $a_{ij}$ are Cartan generators, ladder
operators
and Cartan matrix elements respectively.

This algebra is characterized by Hopf algebra structure; the coproduct
$\Delta$, coidentity $\epsilon$ and antipode $S$ are defined as \cite{F}:
\be \ba{lll}
\Delta H_i &=& H_i\otimes 1+1\otimes H_i \cr
\Delta X_i^{\pm} &=& q^{-H_i/2}\otimes
X_i^{\pm}+X_i^{\pm}\otimes q^{H_i/2} \cr
(\Delta\otimes 1)\Delta X_i^{\pm} &=&
q^{-H_i/2}\otimes q^{-H_i/2}\otimes X_i^{\pm}+q^{-H_i/2}\otimes X_i^{\pm}
\otimes q^{H_i/2} \cr & & +X_i^{\pm}\otimes q^{H_i/2}\otimes
q^{H_i/2},
\ea \ee

\be \ba{lll}
\epsilon(1) = 1  & \hskip1cm \epsilon(H_i)= 0 & \hskip1cm
\epsilon(X_i) = 0 \cr
 S(1)=1     &  \hskip1cm S(H_i)=-H_i  &  \hskip1cm
 S(X_i^{\pm})=-q^{H_i/2}X^{\pm}q^{-H_i/2}.
  \ea \ee
Choosing $a_{ij}$ accordingly, one obtains the standard deformation of
the relevant Lie algebra. Representations of quantum groups can be
constructed by similar routes to the undeformed version.
We are concerned here with the fundamental representations.

For $SU_q(2)$ algebra the matrix representations of $H$ and $X^{\pm}$ are:
\be  \ba{lll}
 &  H = \pmatrix{1&0 \cr 0&-1} & \cr
 X^+ = \pmatrix{0&1 \cr 0&0} & &
X^- = \pmatrix{0&0 \cr 1&0}
\ea    \ee
and for $SU(3)_q$:
\be  \ba{llllll}
H_1 &=& \pmatrix {1&0&0 \cr 0&-1&0 \cr 0&0&0} &
H_2 &=& \pmatrix {1&0&0 \cr 0&1&0 \cr 0&0&-1} \cr \cr
X_1^+ &=& \pmatrix {0&1&0 \cr 0&0&0 \cr 0&0&0} &
X_1^- &=& \pmatrix {0&0&0 \cr 1&0&0 \cr 0&0&0} \cr \cr
X_2^+ &=& \pmatrix {0&0&0 \cr 0&0&1 \cr 0&0&0} &
X_2^- &=& \pmatrix {0&0&0 \cr 0&0&0 \cr 0&1&0}.
\ea \ee
\vskip1cm
Hadrons are made of three quarks each of which
can occur in three types of flavours (up,down,strange),
\be
3 \otimes 3\otimes 3 =3\otimes(6\oplus \bar{3})= 10_S \oplus 8_{MS}
\otimes 8_{MA} \oplus 1_A .
\ee
So there are 27 possible combinations which can be separated into totally
symmetric ($S$), mixed-symmetric ($MS$), mixed-antisymmetric ($MA$) and
totally antisymmetric ($A$) states. By applying the ladder operators
$X_i^{\pm}$ to the highest weight of
these 27 states($uuu$) with the use of the coproduct (2.3), we construct
 the 27 q-deformed states of flavour space. Here we only give an
 example for each $S$, $MS$, $MA$ and $A$ states
  in flavour space.
Now consider the statistics of these particles.
The permutation operator is defined as:
\be \ba{lll}
P_{12} \vert a> \vert b> &=& \vert b> \vert a> \cr
P_{12}^2 &=& 1
\ea \ee
where $\vert a>$, $\vert b>$ are the states of 2 identical particles and
the eigenvalues
are $ \pm 1$,for $S$ and $A$ states.
The eigenstates of $P_{12}$ are:
\be \ba{lll}
\vert ab>^+ &=& \frac{1}{\sqrt{2}}(\vert a> \vert b>+ \vert b>\vert a>) \cr
\vert ab>^- &=& \frac{1}{\sqrt{2}}(\vert a> \vert b>- \vert b>\vert a>).
\ea \ee
The system containing $ N $ identical particles is either totally symmetric
under the interchange of any pair; the eigenvalue of $ P_{12}$ for them is
$+1 $ (Bosons), or totally antisymmetric, with the eigenvalue
$-1$ (Fermions),
\be \ba{lll}
P_{ij} \vert a_1...a_i...a_j...> &=& (+1) \vert a_1...a_j...a_i...> \hskip2cm
for\hskip5mm Bosons \cr
P_{ij} \vert a_1...a_i...a_j...> &=& (-1) \vert a_1...a_j...a_i...> \hskip2cm
for\hskip5mm Fermions.
 \ea \ee
 However all of the q-deformed flavour states are not eigenstates of $P$, in
 fact they are eigenstates of the braid operator $B$.
 To maintain a meaningfull deformation there does not seem to be a choice
 other than letting particle states to be eigenstates of the braid operator.
 This changes our notion of fermions and bosons.

Braids are made of $n$ points on a line that are
 connected by $n$ strings to $n$ point on another parallel line (Fig. 1.a).
  With the operation $b_i$, $i=1,2,...,n-1$,
 we cross over two neighbouring strands $i$ and $i+1$ (Fig. 1.b)
 \cite{Y}.
\begin{figure}
\begin{center}
\setlength{\unitlength}{1mm}
\begin{picture}(140,60)(0,-10)
\multiput(0,5)(0,30){2}{\line(1,0){60}}
\multiput(80,5)(0,30){2}{\line(1,0){60}}

\multiput(0,5)(8,0){2}{\line(0,1){30}}
\multiput(21,5)(8,0){2}{\line(0,1){30}}
\multiput(52,5)(8,0){2}{\line(0,1){30}}
\multiput(80,5)(8,0){2}{\line(0,1){30}}
\multiput(132,5)(8,0){2}{\line(0,1){30}}

\put(-1,1){{\footnotesize $1$}}
\put(7,1){{\footnotesize $2$}}
\put(20,1){{\footnotesize $i$}}
\put(26,1){{\footnotesize $i+1$}}
\put(48,1){{\footnotesize $n-1$}}
\put(59,1){{\footnotesize $n$}}
\put(12,20){\ldots}
\put(38,20){\ldots}

\put(101,5){\line(1,5){6}}
\put(101,35){\line(1,-5){2.6}}
\put(107,5){\line(-1,5){2.6}}

\put(79,1){{\footnotesize $1$}}
\put(87,1){{\footnotesize $2$}}
\put(100,1){{\footnotesize $i$}}
\put(105,1){{\footnotesize $i+1$}}
\put(128,1){{\footnotesize $n-1$}}
\put(139,1){{\footnotesize $n$}}
\put(92,20){\ldots}
\put(118,20){\ldots}
\put(27,-8){{\small (a)}}
\put(107,-8){{\small (b)}}
\end{picture}
\caption{Graphical presentation of $n-braid$ and the operation of
$b_i$ on it.}
\end{center}
\end{figure}
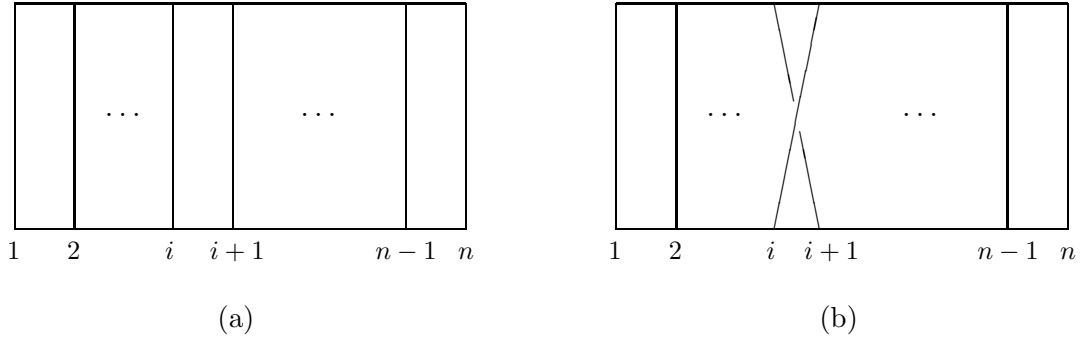
 The simple braid operation satisfies,
 \be \ba{llll}
 b_{i}b_{i+1}b_i&=&b_{i+1}b_ib_{i+1} & \cr
 b_{i}b_j&=&b_jb_i   &  \vert i-j \vert \geq 2.
 \ea  \ee
 The multiplication of braid operators is the placing of braids such
 as Fig. 1.b., thus the identity is the trivial braid of Fig. 1.a.
 The braid group $B_n$, is composed of all the tangling moves possible on
 this structure. In general the group $B_n$ has infinite size, unless
 its generators satisfy some polynomial relationship; the Skein
 relation. The Skein realtion
 offers a way of untangling a braid. For our case, the Skein relation (1.1)
 is equivalent to the  Fig. 2.
With the use of a solution of the Yang-Baxter equation $R$ \cite{B},
 \be
R_{12}R_{13}R_{23}=R_{23}R_{13}R_{12}
\ee
a braid matrix is defined as:
 \be
 B = P^{-1}R
 \ee
where $P$ is the permutation matrix of the equation (2.9).
\begin{figure}
\begin{center}
\setlength{\unitlength}{1mm}
\unitlength=1mm
\begin{picture}(88.00,50.00)

\put(1,35){\line(1,-5){2.6}}

\put(7,5){\line(-1,5){2.6}}

\put(1,5){\line(1,5){6}}

\put(18.00,20.00){\makebox(0,0)[cc]{$- \; q^2$}}
\put(30,35){\line(1,-5){6}}
\put(30,5){\line(1,5){2.6}}
\put(36,35){\line(-1,-5){2.6}}

\put(50.00,20.00){\makebox(0,0)[cc]{$+\;(q^2-1)$}}

\multiput(65,5)(6,0){2}{\line(0,1){30}}
\put(83.00,20.00){\makebox(0,0)[cc]{$= 0$}}
\end{picture}
\caption{The Skein Relation of (1.1)}
\end{center}
\end{figure}
So the matrix of our braid is:
\be  \ba{lll}
B &=& \pmatrix {1&0&0&0\cr 0&1-q^2&q&0\cr 0&q&0&0\cr 0&0&0&1} \cr
\ea \ee
with the eigenvalues $1$ , $-q^2$ for $S$ and $A$ states
respectively,
and now the braid-permutation matrix on a system of two quarks, acts as:
\be \ba{lll}
B \pmatrix{uu \cr ud \cr du \cr dd} &=& \pmatrix {uu \cr (1-q^2)ud+qdu \cr
qud \cr dd} \cr \cr
B \pmatrix{uu \cr us \cr su \cr ss} &=& \pmatrix {uu \cr (1-q^2)us+qsu \cr
qus \cr ss} \cr   \cr
B \pmatrix{dd \cr ds \cr sd \cr ss} &=& \pmatrix {dd \cr (1-q^2)ds+qsd \cr
qds \cr ss}.\cr
\ea \ee
We observe that the exchange of particles is unusual, but
when $q \rightarrow {1}$  we recover the old results.
 Fortunately, the tower of braided states due to the Skein relation (1.1),
does not extend indefinitely.

Now for example, the symmetric state of $\vert \Delta^+>$ under the
exchange of the first two quarks becomes:
\be \ba{lll}
   B_{12} \vert \Delta^+> &=& B_{12} (\frac{1}{\sqrt{1+q^2+q^4}})
   (uud+qudu+q^2duu)
   \cr
 &=&(\frac{1}{\sqrt {1+q^2+q^4}})[uud+q(1-q^2)udu+q^2duu+q^3udu] \cr
 &=& (+1)(\frac{1}{\sqrt {1+q^2+q^4}}) (uud+qudu+q^2duu) \cr
 &=& (+1) \vert \Delta^+>
\ea  \ee
and in the only anti-symmetric state; $\vert \Lambda^o_1>$
 this exchange of the first two quarks  gives:
   \be \ba{lll}
 B_{12} \vert \Lambda^o_1> &=& B_{12}(\frac{1}{\sqrt{(1+q^2)(1+q^2+q^4)}})
 [(sdu-qsud)+(q^2usd-qdsu)+(q^2dus-q^3uds)] \cr
 &=&\frac{1}{\sqrt{(1+q^2)(1+q^2+q^4)}}
 [qdsu-q^2usd+q^2(1-q)-q^2sdu \cr

  & &+q^3uds-q^3(1-q^2)uds-q^4dsu] \cr
 &=&(-q^2)(\frac{1}{\sqrt{(1+q^2)(1+q^2+q^4)}})
 [(sdu-qsud)+(q^2usd-qdsu)+(q^2dus-q^3usd)]
  \cr &=& (-q^2) \vert \Lambda^o_1>.
 \ea \ee
 Therefore, although the state $\vert \Delta^+>$ can be
 interpreted as symmetric, this is not the  case for $\vert \Delta^o_1>$.
 So we are led to require this unorthodox interpretation of permutation of
 identical subsystems of a bound state. This is a fundamental change of
 our notions, but seems inevitable if quantum groups are to find physical
 relevance. This led us to proceed with this proviso in mind
 and look at the physical consequences of this construct.
 Now we see that a hadron state like proton; $\vert p>$,in which the state is
 constructed by a combination of $MS$ and $MA$ states, can not be an
 eigenstate of braid matrix. This problem leads us to conclude that
 the deformation of only one subspace is not enough, that another subspace
 has to be deformed. Spin space seems to be the best choice among
 the others. Tables 1, 2 and 3 show the deformed states for spin and flavour
 states.

 Finally the permutaion matrix is defined as follows:
 \be
 \mbox{\large${\cal P}$}= P_{space} \otimes B_{spin} \otimes B_{flavour}
 \otimes P_{colour}
 \ee
 where the indices define the space on which these operators act. Here
 $B_{spin}$ has the eigenvalues of $1$ and $-q^{-2}$ for the $S$ and $A$
 states respectively. So again we will have $\pm 1$ as the eigenvalues of
 the total permutation matrix. We also note that {\large${\cal P}$}
 is a unitary matrix,
 and an expectation value of an obsevable like $A$ for a particle with the
 state $\vert a>$,
 can have a unitary transfomation under {\large ${\cal P}$}:
 \be
 <a \vert \mbox{\large${\cal P^{\dagger}}$} A \mbox{\large${\cal P}$}\vert a>
 = <a \vert A \vert a>.
 \ee
 \section{Phenomenological Calculations}
 The wave-function of a hadron has to be antisymmetric because,
 quarks are fermions and the wave-function is written as:
\be
\vert \Psi> = \vert Space, Spin> \vert Flavour> \vert Colour>.
\ee
 The
 spape-spin wave function can be separated if the model is non-relativistic
 [15,16].
 It's usual to take the colour state totally antisymmetric,
 so the rest should
be totally  symmetric. For hadrons we have to
consider a combination of $MS$ and $MA$ states
which finally is totally symmetric:
\be
\vert \Psi> = \frac{1}{\sqrt{2}} (\phi_{MS}. \psi_{MS} + \phi_{MA}. \psi_{MA})
\vert Colour>,
\ee
where $\phi$'s and $\psi$'s are the wave functions of space-spin and flavour
respectively.

The static parameters that we have calculated are the magnetic moment $\mu$
\cite{BH}, axial form factor $g_A$ \cite{BH} and Bjorken sum-rule
$g_1^p-g_1^n$ \cite{C}.

The static parameters are:
\be
\ba{lll}
g_A &=& \Delta u - \Delta d  \cr
 & or &   \cr
g_A &=& \sum_{i=1}^3 \sigma_i^3 \tau_i^3\cr
g_1^p-g_1^n &=& \frac{1}{2}( \frac{4}{9} \Delta u + \frac{1}{9} \Delta d)
\ea
\ee
and
\be
 \mu = \sum_{i=1}^3 O_i(x_i,\sigma_i) e_i
 \ee
 where
 \be    \ba{lll}
 < \uparrow \vert O_i \vert \uparrow > &=& a \cr
 < \downarrow \vert O_i \vert \downarrow > &=& -a
   \ea          \ee
in which $i$ sums over the three quarks, $\sigma_i^3$ and $\tau_i^3$ are
the spin  and
   the isospin in the $z$ direction of each of them respectively and
   $\mu$ is the magnetic moment of the hadron with
   "a" as a number depending on the model we use. Here
$\Delta u(d) = u \uparrow(d\uparrow) - u \downarrow (d\downarrow)$
 is the quark asymmetry, $ u \uparrow(\downarrow)$ and
 $ d \uparrow(\downarrow)$ are the probability of finding a u (d) quark
 with spin polarized (unpolarized)to the spin of proton.
 We have calculated these parameters for hadrons by using the wave function
 (3.21) and the states of the tables 1 and 3. In theory,
 for calculating the magnetic moments, it is convenient to find them
 with respect to magnetic moment of neutron.
We have compared the results with the experimental data [17,18] and
found two  Least Square Error(LSE) functions for magnetic moments
and sum rules as follows respectively:
\be    \ba{lll}
LSE(MM)&=&\sum_{i=1}^5 [(\mu_i(q)/\mu(q)_{neutron}-\mu_i^{exp}
/\mu_{neutron}^{exp})/\Delta(\mu_i^{exp}/\mu_{neutron}^{exp})]^2   \cr
LSE(SR)&=&\sum_{i=1}^4 [(g_i(q)-g_i^{exp})/\Delta (g_i^{exp})]^2
\ea \ee
where $i$ sums over the particles and $\Delta$'s are the experimental errors.
For the magnetic moments we have only 5 fuctions because,
the magnetic moment found for $\Xi^o$ is the same
 as of the $n$ and
  the  $ \Sigma^o$  particle is an unstable one.

The experimental results for magnetic moments are more acurate than the sum
rules, therefore we have only used the results of LSE(MM).
The curve is shown at the end of the paper. As it shows, it has two minimums
in $q=0.916,1.0915$.

 Now we give an example of how we found the static parameters in functions
 of $q$, the
 wave function of proton is:
\be \ba{lll}
 \vert P\uparrow > &=& \frac{1}{(1+q^2+q^4) \sqrt{2}}[
 q(1+q^2) u\uparrow d\downarrow u\uparrow -q^4
 u\downarrow d\uparrow u\uparrow - d\uparrow u\downarrow u\uparrow
\cr  & &+ q(1+q^2) d\downarrow u\uparrow u\uparrow- u\uparrow
d\uparrow u\downarrow
-q d\uparrow u\uparrow u\downarrow + q(1+q^2) u\uparrow u\uparrow
d\downarrow  \cr & & -q u\uparrow u\downarrow d\uparrow
-q^3 u\downarrow u\uparrow d\uparrow]
\ea \ee
Here the $\uparrow$,$\downarrow$ stands for spin up or down of the particle.
Next we calculate the expectation value of the magnetic moment operator
giving the following result:
\be
 < P\uparrow \vert \mu \vert P\uparrow > = (\frac{e}{2m}) [\frac{
 7q^2+15q^4+7q^6-q^8-1}{3(1+q^2+q^4)^2}]
\ee
 $m$ is the mass of quark.

The curious thing is that the expectation function of $g_A$ for
the proton is different from neutron,
\be  \ba{lll}
\Delta u &=& \frac{3q^2(1+q^2)^2}{(1+q^2+q^4)^2} \cr
\Delta d &=& \frac{q^8-q^6-3q^4-q^2+1}{(1+q^2+q^4)^2}
\hskip2.5cm all \hskip2mm for \hskip2mm proton \cr
g_A &=& \frac{3q^2(1+q^2)^2-q^8+q^6+q^2+3q^4-1}{(1+q^2+q^4)^2}
 \cr
g_1^p-g_1^n &=& \frac{11q^2+21q^4+11q^6+q^8+1}{18(1+q^2+q^4)^2}
\ea  \ee
and
\be \ba{lll}
\Delta u &=& \frac{q^2-q^4-1}{1+q^2+q^4} \cr
\Delta d &=& \frac{2(1+q^2)}{1+q^2+q^4}
\hskip2.5cm all \hskip2mm for \hskip2mm neutron \cr
g_A &=& \frac{3q^4-q^2+3}{1+q^2+q^4} \cr
g_1^p-g_1^n &=& \frac{7q^4+q^2+7}{18(1+q^2+q^4)},
\ea \ee
and all in $q=1$ give the undeformed results.
\section{Discussions}
The values obtained for q are inverse of eachother, and this shows that
as we have expected,
the algebra is invariant under the exchange of $q \rightarrow q^{-1}$.
{}From the exact symmetry of spin, we also
have  expected that the
values for q should be near 1; $0.9<q<1.1$.
 Note also that although the permutation group was deformed, the concept of
fermions and bosons remains intact at the level of hadrons, if we deform
spin space too.
\\
\\
{\bf Acknowledgments}
\\
The author wishes to thank S. Rouhani, M. Sheikholeslami,
M. Khorrami and F. Arash for their useful discussions.

\newpage
${}$
\begin{table}
\caption{States of q-deformed $1/2$ spin space}
\begin{center}
\begin{tabular}{lr} \hline $States$ & \\ \hline
\\
$S$ & $\alpha(q^2\uparrow \uparrow \downarrow + q\uparrow \downarrow \uparrow
+ \downarrow \uparrow \uparrow)$ \\
$MS$ & $ \gamma[(1+q^2)\uparrow \uparrow \downarrow - (q^3\uparrow
\downarrow \uparrow + q^2 \downarrow \uparrow \uparrow)]$ \\
$MA$ & $ \beta(\uparrow \downarrow \uparrow - q\downarrow \uparrow
\uparrow)$ \\
\end{tabular}
\end{center}
\vskip1cm
\begin{center}
{\small\it Note: $\alpha= \frac{1}{\sqrt{1+q^2+q^4}}$ , $\beta
= \frac{1}{\sqrt{(1+q^2)}}$  and $\gamma=
\frac{1}{\sqrt{(1+q^2)(1+q^2+q^4)}}
$ \\ are the normalization factors.}
\end{center}
\end{table}
\newpage
${}$
\begin{table}
\caption{10 q-symmetric states of flavour subspace}
\begin{center}
\begin{tabular}{lc} \hline
 & $\Phi_S$ \\ \hline
 \\
$\vert \Delta^{++}>$ & uuu \\
$\vert \Delta^{+}>$ & $\alpha(uud+qudu+q^2duu)$ \\
$\vert \Delta^o>$ & $\alpha(udd+qdud+q^2ddu)$ \\
$\vert \Delta^->$ & $ddd$ \\
$\vert \Sigma^{*^{o}}>$ & $\gamma[(usd+q^{-1}uds)+(dus+qsud)+(q^2sdu+qdsu)]$ \\
$\vert \Sigma^{*^{+}}>$ & $\alpha(uus+qusu+q^2suu)$ \\
$\vert \Sigma^{*^{-}}>$ & $\alpha(dds+qdsd+q^2sdd)$ \\
$\vert \Xi^{*^{-}}>$ & $\alpha(dss+qsds+q^2ssd)$ \\
$\vert \Xi^{*^{o}}>$ & $\alpha(uss+qsus+q^2ssu)$ \\
$\vert \Omega^{-}>$ & $sss$

\end{tabular}
\end{center}
\end{table}
${}$
\begin{table}
\caption{8 Mixed q-symmetric and 8
Mixed q-antisymmetric states of flavour subspace}
\begin{center}
\begin{tabular}{lll}  \hline
& $\Phi_{MS}$ & $\Phi_{MA}$ \\ \hline
\\
$\vert p>$ & $\gamma[q(1+q^2)uud-(udu+qduu)]$ & $\beta(qudu-duu)$ \\
$\vert n>$ & $\gamma[(q^2udd+q^3dud)-(1+q^2)ddu]$ &
$\beta(qudd-dud)$ \\
$\vert \Sigma^{+}>$ & $\gamma[q(1+q^2)uus-(usu+qsuu)]$ &
$\beta(qusu-suu)$ \\
$\vert \Sigma^{o}>$ & $\frac{\gamma}{\sqrt{1+q^2}}[(qsud+q^2sdu)
+(usd+qdsu)$
& $\frac{1}{1+q^2}[(q^2dsu+qusd)-(sud+qsdu)]$ \\
 & $-q(1+q^2)(qdus+uds)]$ &  \\
$\vert \Sigma^{-}>$ & $\gamma[q(1+q^2)dds-(dsd+qsdd)]$ &
$\beta(qdsd-sdd)$
\\

$\vert \Lambda^{o}>$ & $\frac{1}{1+q^2}[(-dsu+qusd)
+(q^2sud-qsdu)]$
& $\delta[(q^{-2}sdu-q^{-1}sud)+(usd-q^{-1}dsu)$ \\
& & $-(1+q^2)(dus-quds)]$ \\
$\vert \Xi^{-}>$ & $\gamma[(q^2dss+q^3sds)-(1+q^2)ssd]$ &
$\beta(qdss-sds)$
\\
$\vert \Xi^{o}>$ & $q^{-1}\gamma[(q^2uss+q^3sus)-(1+q^2)ssu]$ &
$\beta(quss-sus)$
\end{tabular}
\end{center}
\vskip 1cm
{\small\it Note: $\delta= \frac{q}{(1+q^2)
\sqrt{q^4+q^2+q^{-2}}}$ is the normalization factors.}
\end{table}


\newpage
\begin{thebibliography}{99}
\bibitem{D}
V.G. Drinfeld " Quantum Groups " , Proc. Internat. Congr. Math. ( Berekley),
vol. 1, Academic Press, New York, (1986)
\bibitem{J}
M. Jimbo, Lett. Math. Phys. 11 (1986) 247
\bibitem{F}
L.D. Fadeev, N.Yu. Reshetikhin, L.A. Takhtajan ; Lenningrad Math. Journal
Vol. 1, (1993) 193
\bibitem{BE}
L. C. Beidenharn, J. Phys. A 28 (1989) L 873
\bibitem{MA}
A. J. Macfarlane, J. Phys. A 28 (1989) 4581
\bibitem{DA}
D. Bonatsos and C. Daskaloyannis, J. Phys. A, Math. Gen. 26 (1993) 1589
\bibitem
S. V. Shabanov, Preprint BUTP-92/24, Institute for Theoretical Physics,
Univ. of Bern
\bibitem{GA}
A. M. Gavrilik and A. V. Tertyohny, ITP Preprint (1993) 19E
\bibitem{BO}
D. Bonatsos et al, J. Phys. 83 (1990) 497
\bibitem{RA}
P. P. Raychev, R. P. Roussev and Yu. F. Smirnov, J. Phys. G 16 (1990) L 137
\bibitem{SH}
M. Sheikholeslami, IPM Preprint
\bibitem{WZ}
W. Schmidke, J. Wess and B. Zumino, Z. Phys. C 52 (1991) 471
\bibitem{B}
R.J. Baxter :"Exactly Solvable Models in Statistical Mechanics", Academic
 Press, London, (1982)
\bibitem{Y}
 Y. Akutsu and M. Wadati, Commun. Math. Phys. 117 (1988) 243-259
\bibitem{C}
F. E. Close :" An Introduction to Quarks and Partons", Academic Press,
London, (1979)
\bibitem{BH}
R. K. Bhaduri :"Models of the Nucleon" , Addison-Wesley Publishing Company,
(1988)
\bibitem{PH}
Phys. Rev. D, Vol. 45, No. 11, 1-June (1992) Part II
\bibitem{PR}
Zein-Eddine Meziani, Proc. of the Summer School on Particle Phys. UC.(1993)
Edited by : L. DePorcel and C. Dunwoodie, page 443-448
\end{thebibliography}
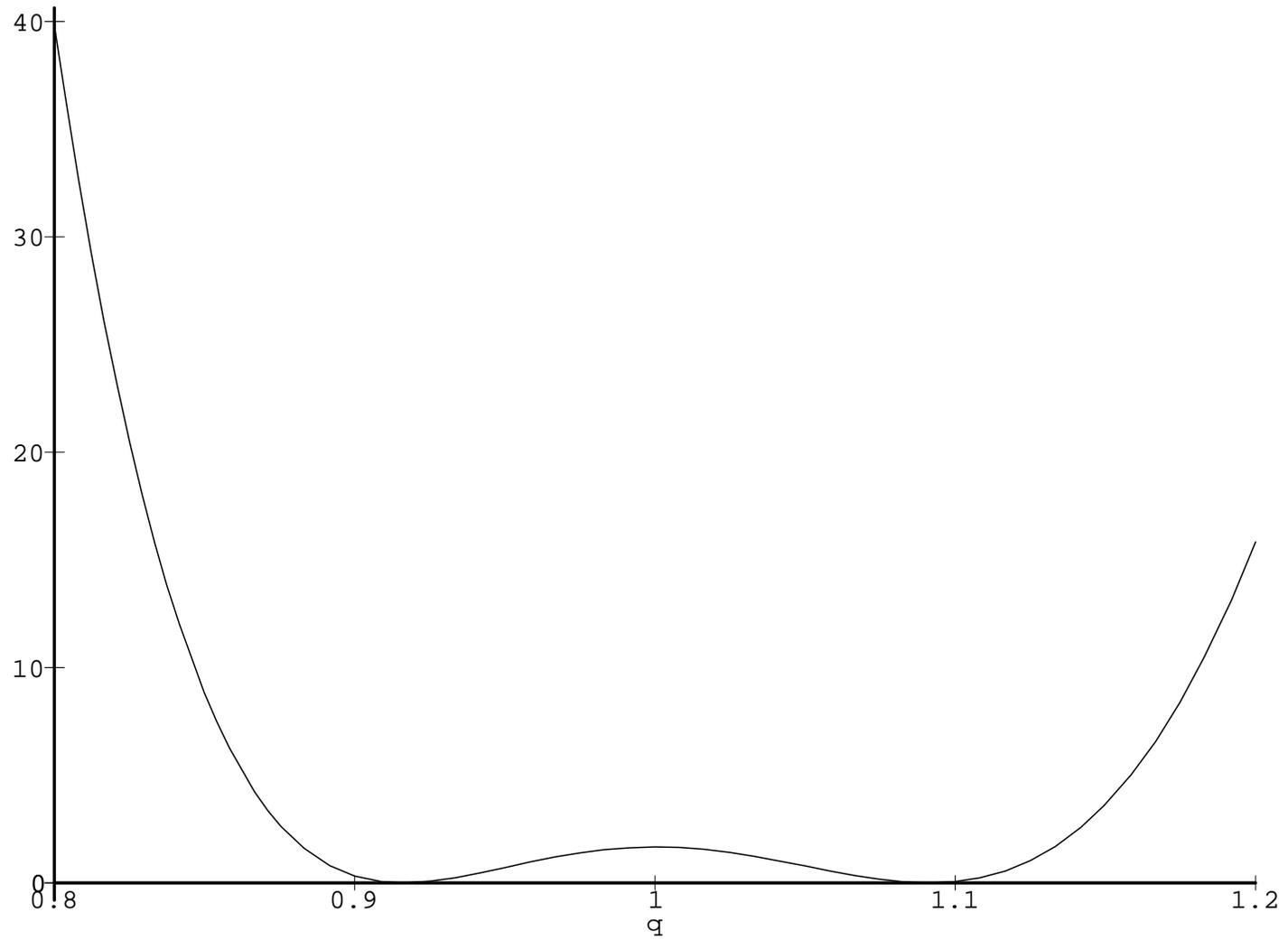
\end{document}